\RequirePackage{silence}
\documentclass[aps,pra,twocolumn,showpacs,amsmath,amssymb,preprintnumbers,superscriptaddress,10pt,longbibliography]{revtex4-2}

\usepackage[bookmarks=false]{hyperref}
\usepackage{graphicx}% Include figure files
\graphicspath{{figures/}}
\usepackage{siunitx}
\usepackage{color}
\usepackage[capitalise]{cleveref}
\crefname{section}{Sec.}{Secs.}% APS style uses abbreviations
\Crefname{section}{Section}{Sections}
\usepackage{xurl}% for proper formatting of URLs in the bibliography
\usepackage{tabularray}
\usepackage{enumitem}

\definecolor{pink}{RGB}{255,0,255}

\definecolor{darkred}{RGB}{140,5,0}

\definecolor{RED}{RGB}{255,0,0}% so that the below also works in section captions that are capitalised
\begin{document}

\title{Wide-spectrum security of quantum key distribution}

\author{Hao~Tan}
\altaffiliation{These authors contributed equally.}
\affiliation{Hefei National Research Center for Physical Sciences at the Microscale and School of Physical Sciences, University of Science and Technology of China, Hefei 230026, People's Republic of China}% Country name is spelled this way per APS style
\affiliation{\mbox{Shanghai Research Center for Quantum Science and CAS Center for Excellence in Quantum Information and} \mbox{Quantum Physics, University of Science and Technology of China, Shanghai 201315, People's Republic of China}}% Country name is spelled this way per APS style
\affiliation{China Telecom Quantum Information Technology Group Co.,\ Ltd.,\ Hefei 230088, People's Republic of China}% Country name is spelled this way per APS style

\author{Mikhail~Petrov}
\altaffiliation{These authors contributed equally.}
\affiliation{Vigo Quantum Communication Center, University of Vigo, Vigo E-36310, Spain}

\author{Weiyang~Zhang}
\affiliation{Hefei National Laboratory, University of Science and Technology of China, Hefei 230088, People's Republic of China}% Country name is spelled this way per APS style

\author{Liying~Han}
\affiliation{Hefei National Research Center for Physical Sciences at the Microscale and School of Physical Sciences, University of Science and Technology of China, Hefei 230026, People's Republic of China}% Country name is spelled this way per APS style
\affiliation{\mbox{Shanghai Research Center for Quantum Science and CAS Center for Excellence in Quantum Information and} \mbox{Quantum Physics, University of Science and Technology of China, Shanghai 201315, People's Republic of China}}% Country name is spelled this way per APS style

\author{Sheng-Kai~Liao}
\affiliation{Hefei National Research Center for Physical Sciences at the Microscale and School of Physical Sciences, University of Science and Technology of China, Hefei 230026, People's Republic of China}% Country name is spelled this way per APS style
\affiliation{\mbox{Shanghai Research Center for Quantum Science and CAS Center for Excellence in Quantum Information and} \mbox{Quantum Physics, University of Science and Technology of China, Shanghai 201315, People's Republic of China}}% Country name is spelled this way per APS style
\affiliation{Hefei National Laboratory, University of Science and Technology of China, Hefei 230088, People's Republic of China}% Country name is spelled this way per APS style

\author{Vadim~Makarov}
\email{makarov@vad1.com}
\affiliation{Vigo Quantum Communication Center, University of Vigo, Vigo E-36310, Spain}
\affiliation{\mbox{Shanghai Research Center for Quantum Science and CAS Center for Excellence in Quantum Information and} \mbox{Quantum Physics, University of Science and Technology of China, Shanghai 201315, People's Republic of China}}% Country name is spelled this way per APS style
	
\author{Feihu~Xu}
\email{feihuxu@ustc.edu.cn}
\affiliation{Hefei National Research Center for Physical Sciences at the Microscale and School of Physical Sciences, University of Science and Technology of China, Hefei 230026, People's Republic of China}% Country name is spelled this way per APS style
\affiliation{\mbox{Shanghai Research Center for Quantum Science and CAS Center for Excellence in Quantum Information and} \mbox{Quantum Physics, University of Science and Technology of China, Shanghai 201315, People's Republic of China}}% Country name is spelled this way per APS style
\affiliation{Hefei National Laboratory, University of Science and Technology of China, Hefei 230088, People's Republic of China}% Country name is spelled this way per APS style

\author{Jian-Wei~Pan}
\affiliation{Hefei National Research Center for Physical Sciences at the Microscale and School of Physical Sciences, University of Science and Technology of China, Hefei 230026, People's Republic of China}% Country name is spelled this way per APS style
\affiliation{\mbox{Shanghai Research Center for Quantum Science and CAS Center for Excellence in Quantum Information and} \mbox{Quantum Physics, University of Science and Technology of China, Shanghai 201315, People's Republic of China}}% Country name is spelled this way per APS style
\affiliation{Hefei National Laboratory, University of Science and Technology of China, Hefei 230088, People's Republic of China}% Country name is spelled this way per APS style

\date{April 29, 2026}
	
\begin{abstract}
Implementations of quantum key distribution (QKD) need vulnerability assessment against loopholes in their optical scheme. Most of the optical attacks involve injecting or receiving extraneous light via the communication channel. An eavesdropper can choose her attack wavelengths arbitrarily within the quantum channel passband to maximise the attack performance, exploiting spectral transparency windows of system components. Here we propose a wide-spectrum security evaluation methodology to achieve full optical spectrum safety for QKD systems. This technique requires transmittance characterisation in a wide spectral band with a high sensitivity. We report a testbench that characterises insertion loss of fiber-optic components in a wide spectral range of 400 to 2300~nm and up to 70~dB dynamic range. To illustrate practical application of the proposed methodology, we give a full Trojan-horse attack analysis for some typical QKD system configurations and discuss briefly induced-photorefraction and detector-backflash attacks. Our methodology can be used for certification of QKD systems.
\end{abstract}

\maketitle

\section{Introduction}
\label{sec:introduction}

Quantum key distribution (QKD)~\cite{bennett1984,ekert1991} allows two remote parties to share secure keys with proven security. Although QKD is proven to be information-theoretically secure, there are gaps between the characteristics of practical devices and their theoretical models~\cite{xu2020}. In such cases, an eavesdropper Eve might perform attacks and eavesdrop the secure key, compromising the security of QKD systems \cite{diamanti2016}. In QKD's structure, the communicating parties Alice and Bob reside in physically secure locations, and Eve cannot access them directly. However, the communication channel is controlled by Eve, and she can inject light into Alice's and Bob's apparatuses through this channel to affect the state of internal components and also extract light reflected or emitted from them. Thus some of the most threatening strategies are light-injection and unintended-light-emission attacks \cite{vakhitov2001,gisin2006,makarov2006,makarov2009,xu2010,lydersen2010a,li2011a,jain2014,sajeed2015,bugge2014,sun2015,shi2017,pinheiro2018,huang2019,wei2019,huang2020,tan2021,ponosova2022,ye2023,han2023}.

For the transmitters of QKD systems, Eve can inject light in order to eavesdrop modulator's encoding information via reflected light \cite{vakhitov2001,gisin2006,jain2014}, alter the performance of the laser \cite{sun2015,huang2019,fadeev2025}, damage the optical devices \cite{huang2020,ponosova2022}, change the operating characteristics of phase and intensity modulators \cite{ye2023,han2023} and energy meters \cite{sajeed2015}. For the receivers of QKD systems, Eve can inject light to control the detectors \cite{makarov2006,makarov2009,lydersen2010a}, damage them \cite{bugge2014}, eavesdrop modulator's encoding information \cite{jain2014,sajeed2017}, or receive light emission from avalanche photon detectors leaked back into the communication channel \cite{meda2017,shi2017,pinheiro2018}.

Normally, QKD systems need to deploy additional passive optical devices, such as isolators, attenuators, and spectral filters, to mitigate the effects of these attacks \cite{lucamarini2015,tan2021,ponosova2022}. However, the transmittance of these passive optical devices is typically specified only around their design wavelength, and shows significant deviations at other wavelengths, which may lead to potential spectral side-channels. The communication channel itself, either free-space or optical fiber, is transparent in a very wide spectral range. Eve can thus choose her working wavelength arbitrarily to maximise the attack performance. For example, the reverse transmittance of isolators and circulators designed for 1550~\si{\nano\meter} rises significantly in the 1000--1400~\si{\nano\meter} wavelength range \cite{jain2015}. This spectral side-channel may provide a valuable ``window of attack'' for eavesdroppers and compromise the security of QKD systems. In another example, a Trojan-horse attack on a 1550-\si{\nano\meter} QKD receiver becomes feasible at 1924~\si{\nano\meter} \cite{sajeed2017}. An avalanche backflash is spectrally broad, spanning hundreds of nanometers \cite{shi2017,pinheiro2018}. Recently, an induced-photorefraction attack using light at 532~\si{\nano\meter} \cite{han2023} and 405~\si{\nano\meter} \cite{ye2023} has been investigated, which results in significant changes in the refractive index and transmittance of lithium-niobate phase and intensity modulators. Therefore, we need to carefully analyse the vulnerability of optical components in a wide spectral range and propose appropriate countermeasures accordingly to guarantee the full-spectrum resistance to attacks in QKD systems. This will be necessary for upcoming certification of QKD \cite{iso23837-2023,makarov2024,tomita2019,sajeed2021,a-etsi2021}. This applies to both discrete-variable and continuous-variable systems, as the latter are also susceptible to Trojan-horse, induced-photorefraction, and other wavelength-dependent attacks \cite{zhang2024,iso23837-2023,marquardt2023}.

Here we propose a characterisation methodology to achieve this. We consider a general approach for protection of cryptographic modules and classify attacks into three cases (see~\cref{sec:theory}). We build an experimental testbench that can characterise transmittance of fiber-optic components in a wide spectral range of 400 to 2300~\si{\nano\meter}. We supplement the cryptographic module with a bandpass physical filter that guarantees a strong suppression of all light outside this characterisation window. This allows us to find the weakest spectral spot in the protection of a QKD scheme against each attack. The information leakage is quantified at this wavelength, then if necessary the protection is improved (by, e.g.,\ installing additional isolating components or changing the optical scheme) to a point where the attack does not impair QKD performance.

We demonstrate our methodology on examples of the Trojan-horse attack (THA), induced-photorefraction, and detector-backflash attacks. For the former, we consider three QKD source configurations and characterise each optical component in these configurations using our testbench. The optical components include isolators, circulators, variable optical attenuators, filters, and fiber coils. These devices exhibit different transmission characteristics across the wavelength spectrum. Based on these measurements, we analyse the secure key rate under THA for two QKD schemes---prepare-and-measure BB84 and measurement-device-independent---using existing security proofs and identify the source configuration that makes each scheme fully resistant to this attack. We then consider briefly the two other types of attacks, induced-photorefraction and detector-backflash.

This paper is organised as follows. We discuss the wide-spectrum security evaluation methodology in \cref{sec:theory}. \Cref{sec:setup} introduces the testbench hardware and test methodology. In \Cref{sec:physical_filter}, we discuss how to design the bandpass physical filter. We give application examples in \cref{sec:evaluation} and conclude in \cref{sec:conclusion}.

\section{Methodology}
\label{sec:theory}

Eavesdropping attacks on a QKD system involve either Eve's light injection via a quantum channel to affect a specific internal component or analysis of the system's unintended light emission. In the first scenario, Eve controls an individual component or obtains information about it via reflected light. In the second scenario she doesn't inject light into the system, but only registers and analyses the light emitted. In both scenarios, the transparency of her attack channel (i.e., that of components between the target and the quantum channel) is essential. If it is attenuating light too much, Eve is unable to affect the target component and can't collect enough photons coming out of the system that contain secret information. Typically, QKD vulnerability is assessed only at the operating wavelength (most common in C-band), but it is essential to consider how transparent the attack channel is throughout the entire spectral range available via the quantum channel. A spectral transparency window allows Eve to more actively target QKD internal components or get a more intense light signal from them, and as a result, steal more information. For each possible attack scenario, the QKD vulnerability analysis involves identifying the optimal wavelength (from Eve's viewpoint) corresponding to the maximum level of component interference or light leakage, and subsequent estimate of the maximum information leakage from the QKD devices. The analysis for a particular QKD implementation consists of several steps.
\begin{itemize}[label=$\circ$,leftmargin=2ex,noitemsep]
\item The QKD implementation is analyzed for the feasibility of all currently known attacks (similarly to \cite{makarov2024,sajeed2021}).  
\item For each possible attack, the attack channel and list of its constituent components are defined.
\item The transmission of each of these components is characterised over a wide spectrum.
\item The total transmission spectrum of the attack channel is calculated as a product of the individual component transmittances.
\item The response of the target component to illumination at different wavelengths or its emission spectrum is characterised.
\item The maximum power of Eve's light at the target component at the optimal wavelength for attack or total power exiting the QKD system is estimated.
\item From this estimate, amount of information leakage is evaluated using existing security proofs or other experimental studies.
\end{itemize}

This wide-spectrum methodology is applicable for a very broad class of attacks on QKD systems with different protocols and internal designs. The vulnerability analysis for each specific QKD implementation should be performed individually. Every system is unique and susceptible to its own specific set of attacks \cite{sajeed2021,marquardt2023,iso23837-2023,makarov2024}. Each has its own set of components, moreover the properties of components of the same type differ from one manufacturer to another.

More formally, the methodology is the following. To evaluate security against a light-injection or light-emission attack, one needs to know the spectral response of its target component to the attack light $S (\lambda)$ and the total transmittance of an attack channel that the light traverses inside the cryptography equipment during the attack $\gamma (\lambda)$. Both parameters are wavelength-dependent. Then, a secure key rate $R = K[S(\lambda),\gamma(\lambda)]$ is calculated. Our study is primarily concerned with the measurement of $\gamma (\lambda)$, which can be done in a uniform way. The measurement of $S (\lambda)$ and derivation of the function $K$ are attack-specific, though we discuss some examples of these below.

\begin{figure}
	\includegraphics{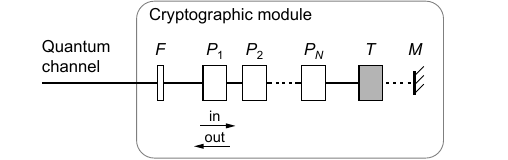}
	\caption{General structure of a cryptographic module from Eve's point of view. An optical component $T$ is the target of attack. It is separated from the quantum channel by other optical components $P_1$ to $P_N$ and a broadband bandpass physical filter $F$ that blocks very short and very long wavelengths. For some attacks, a backreflection $M$ behind $T$ should be taken into account.}
	\label{fig:configuration}
\end{figure}

From Eve's perspective, a secure cryptographic module (such as Alice or Bob) has a structure shown in \cref{fig:configuration}. The target component of Eve's attack $T$ is connected to the quantum channel, where Eve resides, via optical components $P_1$ to $P_N$ and a filter $F$ (which we explain later). Behind $T$, non-zero backreflections from connectors and other components always exist \cite{vakhitov2001}, collectively denoted by a reflection coefficient $M$.

An important parameter for Eve's attack is the total transmittance of the attack channel $\gamma(\lambda)$. The light suffers attenuation as it passes every component on its way. We classify Eve's attacks into three cases.

\emph{Case~1}. For light-injection attacks such as laser-seeding, laser-pumping, induced-photorefraction, laser-damage, detector-efficiency-mismatch, and tampering with a power meter, the attack channel is unidirectional from Eve to $T$. Then
\begin{equation}
	\label{eq:configuration-in}
	\gamma(\lambda) = F(\lambda) \prod_{i=1}^N P_i^\text{in}(\lambda),
\end{equation}
where $P_i^\text{in}$ denotes a transmittance of component in the inward light propagation direction. While many optical components tend to have the same transmittance in both directions, some may be slightly direction-dependent or even be designed for strong non-reciprocity, such as optical isolators and circulators. For simplicity, we assume the filter's transmittance $F$ is the same in both directions.

\emph{Case~2}. Another case is a leakage of photons generated inside the cryptographic module into the communication channel, of which we currently know one example, the detector-backflash attack. In this case, the attack channel is unidirectional from $T$ to Eve. Its total transmittance
\begin{equation}
	\label{eq:configuration-out}
	\gamma(\lambda) = F(\lambda) \prod_{i=1}^N P_i^\text{out}(\lambda),
\end{equation}
where $P_i^\text{out}$ is the component's transmittance in the outward direction. In these two cases, reflection $M$ is neglected.

\emph{Case~3}. The third case is the Trojan-horse attack where Eve injects light and receives a fraction of it reflected back, to learn the state of the modulator. The attack channel is thus a round-trip. Then
\begin{equation}
	\label{eq:configuration-in-out}
	\gamma(\lambda) = \left[F(\lambda)\right]^2 T^\text{in}(\lambda) T^\text{out}(\lambda) M(\lambda) \prod_{i=1}^N P_i^\text{in}(\lambda) P_i^\text{out}(\lambda),
\end{equation}
where $T^\text{in}$ and $T^\text{out}$ is the transmittance of the target modulator in either direction. Estimating a reliable upper bound on $M$ is experimentally difficult, owing to the possibility of constructive interference between multiple reflections Eve might exploit, high temporal resolution required, and possible dependence of reflections on wavelength \cite{a-etsi2021}. Also, an internal reflection inside the target modulator behind its active section may contribute, excluding part of the modulator's attenuation. To avoid these challenges, it is much easier to conservatively assume $T^\text{in} T^\text{out} M = 1$ \cite{a-etsi2021}. Then \cref{eq:configuration-in-out} simplifies to
\begin{equation}
	\label{eq:configuration-in-out-simplified}
	\gamma(\lambda) = \left[F(\lambda)\right]^2 \prod_{i=1}^N P_i^\text{in}(\lambda) P_i^\text{out}(\lambda).
\end{equation}

Since $\gamma(\lambda)$ is a product of individual transmittances (i.e.,\ any possible interference effect involving reflections is neglected), it can be computed from transmittances of the individual optical components. We measure these with our testbench reported below, which covers a wide but finite spectral range. The transmittance of components remains unknown outside this characterisation range. We thus need to supplement each cryptographic module with the filter $F$ that guarantees light suppression outside this range by its design (discussed in \cref{sec:physical_filter}).

The susceptibility of the target component itself to the attack light $S(\lambda)$ also varies with wavelength. While for treatment of the Trojan-horse attack it may reasonably (and conservatively) be assumed to be spectrally flat \cite{lucamarini2015}, for most attacks it varies strongly with wavelength \cite{sun2015,huang2019,sajeed2017,ye2023,han2023}. For instance, Alice's laser is much more responsive to seeding by light near its emission wavelength \cite{sun2015,huang2019} and Bob's avalanche photodiode emits its backflash in a broad but limited spectral band \cite{meda2017,shi2017,pinheiro2018}. Characterising this susceptibility and attack mechanisms is outside the scope of our present study.

Finally, the form of function $K$ also depends on the attack. Often, Eve's best strategy is to exploit the maximum response $\max_{\lambda}[\gamma(\lambda) S(\lambda)]$ and inject light at that single wavelength, such as, for example, in the Trojan-horse \cite{lucamarini2015} and induced-photorefraction \cite{ye2023,han2023} attacks. However, some attacks may be best executed at multiple discrete wavelengths or over a certain spectrum. For instance, in the detector-backflash attack, the probability of a photon leaking into the quantum channel should be integrated over the entire emission spectrum $\int_\lambda \gamma(\lambda) S(\lambda) \,d\lambda$, where $S(\lambda)$ is the measured probability density of photon emission from the avalanche photodetector. The key rate formula $K$ is further given by a relevant security proof that accounts for the attack or for several attacks simultaneously. In \cref{sec:evaluation}, we give some examples of this analysis for particular attacks.

\section{Spectral characterisation testbench}
\label{sec:setup}

\begin{figure}
	\includegraphics{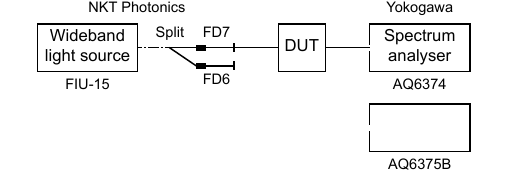}
	\caption{Scheme of the spectral characterisation testbench. DUT, device under test.}
	\label{fig:setup}
\end{figure}

Manufacturers of components typically do not document their spectral characteristics over the entire transparency range of the quantum channel that can be used by Eve. We thus have to measure the component transmittance ourselves. This measurement should have a wide dynamic range, because the component may have a low transmittance. Here we implement the testbench proposed in \cite{makarov2024}. It consists of a wideband light source, fiber coupler, and spectrum analyser (\cref{fig:setup}). We use a supercontinuum laser source (NKT Photonics SuperK Fianium FIU-15) that emits ``white light'' of about $350$--$2400$~\si{\nano\meter} spectrum with variable pulse repetition rate of $15$--$78~\si{\MHz}$ and total power of up to $7$~\si{\watt} \cite{NKT-fiu-15}. Compared to a previous testbench \cite{nasedkin2023}, the spectral coverage of our testbench is wider, facilitating the analysis of attacks in specific bands, such as the induced-photorefraction attack \cite{ye2023,han2023}. The power can be trimmed by lowering the pulse rate. Although its output is already single-mode in a large-core-diameter fiber, it needs to be coupled into the standard single-mode fiber used by the components we test. The fiber coupler consists of a dichroic splitter (NKT Photonics SuperK Split \cite{NKT-split}) that separates the spectrum into two $400$--$900~\si{\nano\meter}$ and $900$--$2400~\si{\nano\meter}$ outputs, each fitted with a tunable fiber coupler (NKT Photonics SuperK Connect FD7 and FD6, respectively \cite{NKT-connect}). The testbench operator manually connects the device under test (DUT) to these outputs sequentially to scan the entire wavelength range. Likewise, two spectrum analysers are used interchangeably to cover the entire wavelength range: Yokogawa AQ6374 ($350$--$1750$~\si{\nano\meter} \cite{Yokogawa-AQ6374}) and AQ6375B ($1200$--$2400$~\si{\nano\meter} \cite{Yokogawa-AQ6375B}).

We limit the total power at the input of DUT to $0.5$~\si{\watt}, to avoid damage to fiber patchcords and the spectrum analysers. The power is lowered further for temperature-sensitive DUTs, such as variable attenuators and narrow spectral filters. Incidentally, the dichroic splitting into the two bands helps to suppress ghosts due to higher-order diffraction in the spectrometers. We limit our total characterisation range to $400$--$2300~\si{\nano\meter}$, because below $400~\si{\nano\meter}$ the light source has low spectral flux and the spectrometer has higher noise, and above $2300~\si{\nano\meter}$ the suppression of the ghost in the spectrometer is poor. To overcome the limitation of characterisation wavelength range and ensure security beyond it, we design physical filters to limit Eve's attack wavelengths, which will be discussed in \cref{sec:physical_filter}.

\begin{figure*}
	\includegraphics{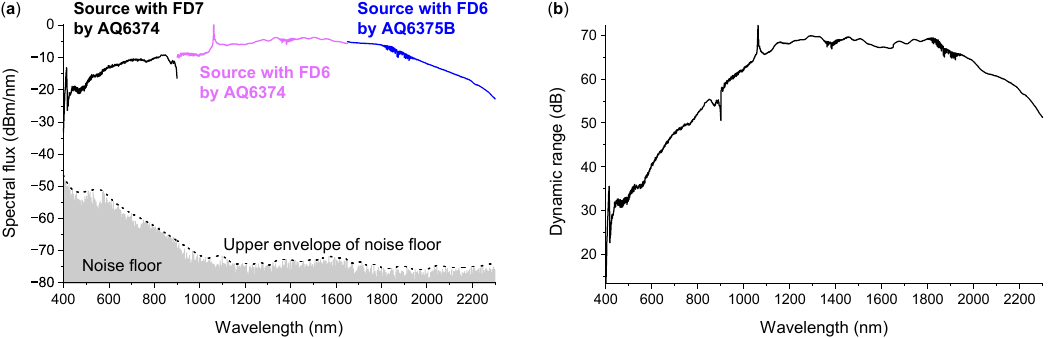}
	\caption{Spectral performance of the testbench. (a)~Spectrum of the light source (measured with the DUT replaced with a patchcord) and the analyser's dark noise, at 1~\si{\nano\meter} resolution. (b)~The resulting dynamic range of the transmittance measurement. This data is at about $0.5$~\si{\watt} at the DUT; the dynamic range will be narrower at a lower power.}
	\label{fig:laser_source}
\end{figure*}

Obtaining the component's transmittance consists of a scan of the light source (done with the DUT replaced with a patchcord) followed by the same scan with the DUT in place. We have verified that the source and the rest of the setup remain stable between these scans. The DUT's transmittance is then a ratio of the spectral flux measured with the DUT and without the DUT. In logarithmic units of spectral flux, the ratio is replaced with subtraction. The entire spectrum is covered by repeating these scans in three setup configurations and stitching them together, with the stitching point between the splitter's outputs being at $900~\si{\nano\meter}$ and between the two spectrometers at $1650~\si{\nano\meter}$. A typical source scan is shown in \cref{fig:laser_source}(a). It also shows the spectrometer dark noise level and its smoothed upper envelope obtained by a local maximum method combined with a cubic spline interpolation (function ``Envelope'' in OriginPro software). The difference between the measured source flux and the noise envelope gives the dynamic range of our transmittance measurement, plotted in \cref{fig:laser_source}(b). It is above $65$~\si{\decibel} in the central part ($1100$--$1900$~\si{\nano\meter}) and drops to 30--50~\si{\decibel} at the edges. When the component's insertion loss at a given wavelength is higher that that, we are unable to measure its transmittance and have to replace it with a value given by the noise envelope, for security evaluation purposes. We are thus able to guarantee the reliability of the measurement data. Although this approach is conservative, it gives adequate results when applied to typical QKD configurations, as the reader will see in \cref{sec:evaluation}.

The spectral characterisation is done with 1~\si{\nano\meter} resolution, which is sufficient for most components. If the component contains a Bragg grating filter or other structure that produces sharp spectral features, the scan resolution can be increased to 0.05~\si{\nano\meter}, at the cost of a lower dynamic range.

\section{Design of a broadband bandpass physical filter}
\label{sec:physical_filter}

The standard single-mode fiber may transmit light outside the characterisation wavelength range of our testbench. Since the behaviour of the system components there remains unknown, a supplementary filter $F$ is needed. This filter should guarantee suppression of short- and long-wavelength light to a secure level even if all the other system components are conservatively assumed to be transparent. This completes the system certification against the attack. The suppression has to be provided by the physical design of the filter rather than by our measurement.

\begin{figure}
	\includegraphics{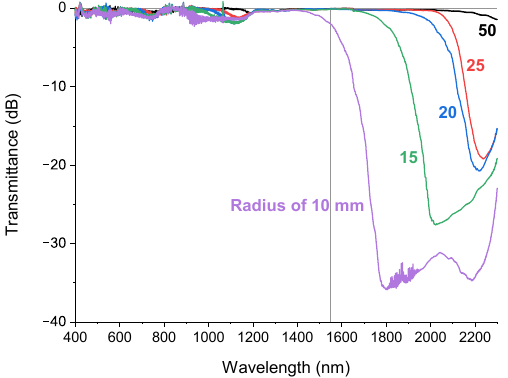}
	\caption{Transmittance of a 2-\si{\meter} long single-mode fiber patchcord coiled for its entire length with different radii. The fiber is Corning SMF-28e+ with $242~\si{\micro\meter}$ diameter coating inside a 0.9-\si{\milli\meter} outer diameter loose jacket, made into the patchcord with FC/UPC connectors (Optizone Technology P-55-R-11-L-F-2).}
	\label{fig:fiber_coils}
\end{figure}

For long-wavelength suppression, bend loss of the fiber may be used. If the fiber is coiled with a sufficiently small radius, it loses its waveguiding properties at longer wavelengths, releasing light into the cladding where it is emitted or absorbed in the coating~\cite{wang2005c}. The smaller the bending radius, the higher the loss of the fiber. Furthermore, light with long wavelength is more easily leaked out of the fiber at a specific radius. For example, a standard single-mode fiber exhibits high loss at longer wavelengths, see our measurement in \cref{fig:fiber_coils}. The effective cutoff wavelength depends on the bend radius, with 15~\si{\milli\meter} radius being suitable for a typical QKD system. This broadly agrees with the theory, which predicts a significant increase in bend loss at 1550~\si{\nano\meter} when the radius is less than 12~\si{\milli\meter} \cite{wang2005c}. A previous experiment \cite{chen2019} shows that a single-loop coil of 15-\si{\milli\meter} radius has about 30~\si{\decibel} attenuation at 2250~\si{\nano\meter} and virtually no attenuation at 1550~\si{\nano\meter}, which also roughly agrees with our result. The high residual transmission observed in \cref{fig:fiber_coils} at longer wavelengths is probably an artefact of our quick measurement, and should be verified in a more carefully controlled test with a different turn number and the coil being connected in series with other system components to filter out cladding modes.

For short-wavelength suppression, well-understood optical properties of bulk material may be used. For example, pure Si has an absorption coefficient $\alpha \gtrsim 10^{5}~\si{\centi\meter}^{-1}$ in $250$--$400~\si{\nano\meter}$ range, while being virtually transparent ($\alpha<1~\si{\centi\meter}^{-1}$) past $1140~\si{\nano\meter}$ \cite{green1995, green2008}. A filter can be manufactured inexpensively by inserting a several-millimeter-thick silicon optical window into a fiber-pigtailed collimator bench. Its cutoff may be shifted to a shorter wavelength by selecting a semiconductor material with a wider bandgap than that of Si ($1.12~\si{\electronvolt}$ \cite{kasap2006}). For example, GaAs (GaP) with the bandgap of $1.42$ ($2.26$)~$\si{\electronvolt}$ \cite{kasap2006} has the cutoff wavelength around $870$ ($550$)~$\si{\nano\meter}$.

The design of these filters should be finalised together with the system manufacturer and further verified by tests and theoretical calculations. To protect $F$ against laser damage, it may have to be placed not at the channel entrance as shown in \cref{fig:configuration}, but behind other components \cite{ponosova2022}. This is future work.

\section{Application examples}
\label{sec:evaluation}

While our wide-spectrum methodology can be used for vulnerability assessments for every known attack that depends on light injection or emission, we give a few examples of its use. First, we give a full analysis (complete with the key rate calculation) of the Trojan-horse attack for some typical QKD system configurations. Then we discuss briefly how to approach two other types of attacks, induced-photorefraction and detector-backflash.

\subsection{Trojan-horse attack}

In the Trojan-horse attack, Eve sends bright light into Alice or Bob, where it passes through their modulators and gets partially reflected back into the communication channel \cite{vakhitov2001,gisin2006,jain2014,jain2015,sajeed2017}. She can then measure this reflection and learn the state of the modulator surreptitiously, without disturbing QKD operation. In fiber-optic QKD systems, the intensity of Eve's injected light is upper-bounded by a laser-induced damage threshold of the single-mode fiber comprising the channel. Given a sufficiently low $\gamma (\lambda)$, Eve receives much less than one reflected photon per qubit and the key information leakage is partial \cite{lucamarini2015}. It is determined by the mean reflected photon number $\mu_\text{out} = N \gamma / f$, where $N$ is the total number of photons per second Eve may inject into the channel and $f$ is the clock rate of the QKD system. 

\begin{figure}
	\includegraphics{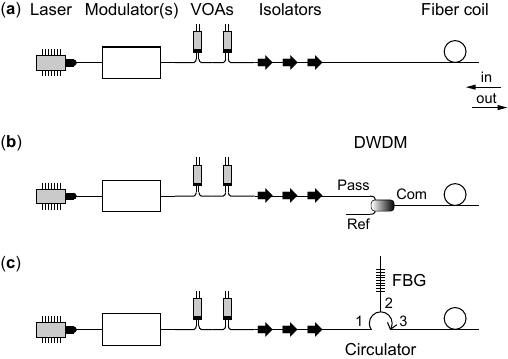}
	\caption{Source configurations. The fiber coil is used as the broadband physical filter. The passive protection components in configuration (a) are VOAs and isolators. In configuration (b), we add a dense wavelength division multiplexer (DWDM). In configuration (c), the DWDM is replaced with a fiber Bragg grating (FBG) filter. Arrows indicate the direction of attack light transmission. Arrows in the isolator and circulator symbols indicate their forward transmission direction.}
	\label{fig:source_configurations}
\end{figure}

The source in the QKD system often consists of a laser followed by modulator(s) that prepare different quantum states and variable attenuators (VOAs) that attenuate them to the required intensity. For protection against the THA, optical isolators and filters are added to the source, in order to decrease $\gamma (\lambda)$. Here we investigate three source configurations (\cref{fig:source_configurations}), which were proposed to mitigate the impact of THA \cite{lucamarini2015} and are widely deployed in commercial QKD devices \cite{makarov2024}. All of them use two VOAs, three isolators, and the fiber coil of 15~\si{\milli\meter} radius and 21 turns. Configuration (a) consists of only these components. The isolator transmits light at the laser wavelength in the forward direction and heavily attenuates it in the reverse direction.

In configuration (b), a dense wavelength division multiplexer (DWDM) is added. It is a three-port device that internally consists of fiber collimators and a thin-film filter. It is designed to transmit light between the ports $\text{Com} \rightleftarrows \text{Pass}$ near the laser wavelength and $\text{Com} \rightleftarrows \text{Ref}$ at other wavelengths within its working wavelength range.

In configuration (c), the DWDM is replaced with a filter consisting of a fiber Bragg grating (FBG) and a circulator. The fiber Bragg grating is a diffraction grating with a periodic change of refractive index in the fiber core. It transmits most light without attenuation but strongly reflects back the laser light whose wavelength matches two grating periods. The circulator transmits light from its port 1 to 2 and from port 2 to 3, and heavily attenuates light going in all the other directions. This filter thus passes laser light to the source output and heavily attenuates all other light in both directions.

In order to compute $\gamma (\lambda)$ via \cref{eq:configuration-in-out-simplified}, we measure transmittance of individual components using our testbench.

\subsubsection{Characterisation of individual components}	
\label{sec:components}

\begin{figure}
	\includegraphics{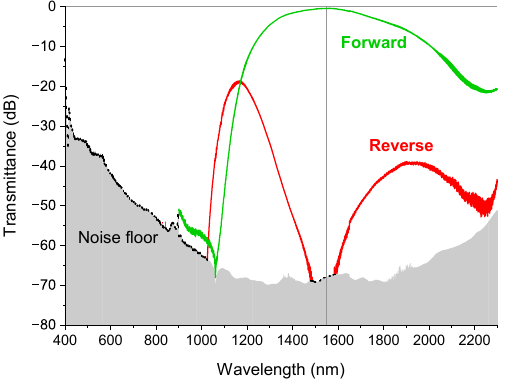}
	\caption{Spectral characteristics of the isolator. The spectrometer noise envelope is shown as a gray area (omitted for clarity in subsequent plots) and as a black dotted line wherever it substitutes real transmittance data.}
	\label{fig:ISO}
\end{figure}

\emph{Isolator.} We test a polarisation-insensitive isolator (Optizone Technology PII-55-P-T-2-11-LL-1). Its transmittance in both forward and reverse directions is shown in \cref{fig:ISO}. The sensitivity of our spectrometer did not allow us to measure the true transmittance in parts of the spectral range, and the noise envelope was substituted there. The forward transmittance is about $-0.5~\si{\decibel}$ near the working wavelength of $1550~\si{\nano\meter}$ but drops far from it. The isolation (defined by the reverse transmittance) is high around $1550~\si{\nano\meter}$ but deteriorates far from it, especially in the $1050$--$1300~\si{\nano\meter}$ range. There it drops at least $50~\si{\decibel}$ lower than at the working wavelength.

\begin{figure}
	\includegraphics{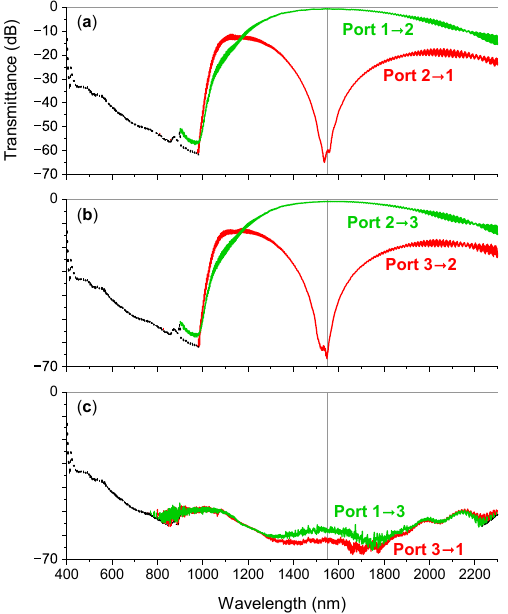}
	\caption{Spectral characteristics of the circulator between ports (a)~1 and 2, (b)~2 and 3, and (c)~1 and 3.}
	\label{fig:CIR}
\end{figure}

\medskip
\emph{Circulator.} We test a polarisation-insensitive circulator (Optizone Technology FCIR-55-1-111-LLL-1) used in our source configuration~(c). Its transmittance in all six possible directions is shown in \cref{fig:CIR}. While the characteristics between ports $1 \rightleftarrows 2$ and $2 \rightleftarrows 3$ are broadly similar to those of the isolator, the attenuation between ports $1 \rightleftarrows 3$ remains uniformly high at all wavelengths.

\begin{figure}
	\includegraphics{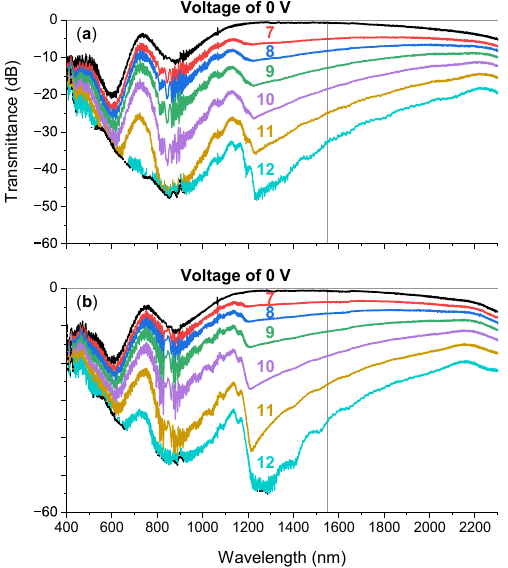}
	\caption{Spectral characteristics of MEMS VOA at different control voltages. (a)~Sample~1; (b)~sample~2.}
	\label{fig:VOA}
\end{figure}

\medskip
\emph{Variable optical attenuator.} Industrial QKD systems often use inexpensive micro-electro-mechanical systems (MEMS) VOAs. We test two samples of one of these devices (Shanghai Honghui optics communication material Co.,\ Ltd.\ FASRE-55SM-40BR-LS-FU-1M). It contains a mirror deflected by an externally applied dc voltage, changing its attenuation in about $1$--$30~\si{\decibel}$ range at $1550~\si{\nano\meter}$. However, as can be seen in \cref{fig:VOA}, the attenuation at longer wavelengths can be significantly lower than that set.

\begin{figure}
	\includegraphics{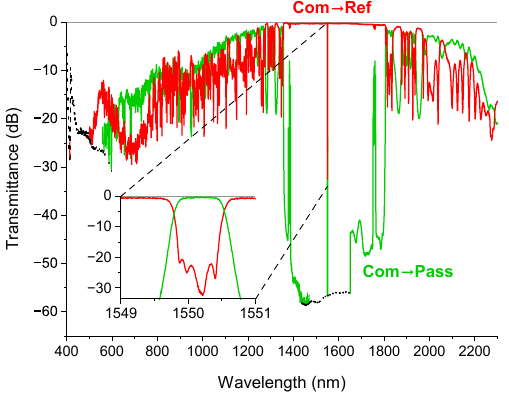}
	\caption{Spectral characteristics of DWDM. The scan is at 1-\si{\nano\meter} resolution except its 1549--1551~\si{\nano\meter} portion scanned at 0.05~\si{\nano\meter} resolution (magnified in the inset).}
	\label{fig:DWDM}
\end{figure}

\medskip
\emph{Dense wavelength division multiplexer.} Quantum key distribution systems sometimes employ the DWDM to add separate synchronisation and data channels at different wavelengths. We test one such device (Connet Fiber Optics DWDM-100G-1$\times$2-C34-900-1-FA), see \cref{fig:DWDM}. Only one direction is shown; transmission in the opposite direction ($\text{Ref} \rightarrow \text{Com}$ and $\text{Pass} \rightarrow \text{Com}$) is virtually identical. The device is working as designed in the $1400$--$1650~\si{\nano\meter}$ range. Outside this range, its transmission through both port pairs becomes chaotic and generally high.

\begin{figure}
	\includegraphics{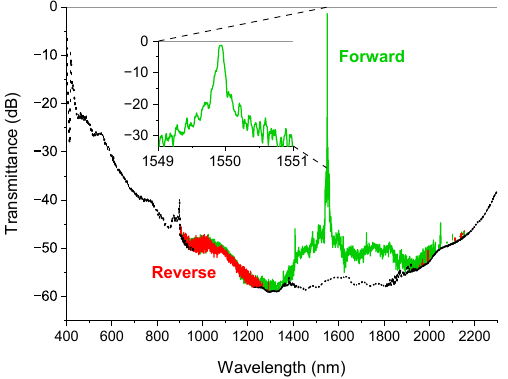}
	\caption{Spectral characteristics of the FBG-based filter. The scan is at 1-\si{\nano\meter} resolution except its 1549--1551~\si{\nano\meter} portion scanned at 0.05~\si{\nano\meter} resolution (magnified in the inset).}
	\label{fig:FBG}
\end{figure}

\medskip
\emph{Fiber Bragg grating-based filter.} The FBG filter consists of the circulator tested above and FBG (Wuxi Ruike Huatai Electronics Limited FBG-100) connected at its port~2. The other end of FBG is pigtailed with an angled connector that remains unconnected. We define the transmission from the circulator's port 1 to port 3 as the forward direction and the opposite as the reverse direction. We test this filter as a whole, see \cref{fig:FBG}. Its attenuation is above $40~\si{\decibel}$ through the $750$--$2270~\si{\nano\meter}$ range in both directions, with a single narrow forward transmission peak at $1550~\si{\nano\meter}$. This filter is thus wideband, unlike the other components.

\subsubsection{Security evaluation}
\label{sec:security}

The impact of information leakage from the encoding module caused by the THA needs to be quantified. Since the target susceptibility $S$ to THA is assumed to be wavelength-independent and need not be tested, the above measurements are sufficient for protocol security analysis \cite{lucamarini2015}. The targets of the THA are the phase and intensity modulators. The former may leak the basis information under the THA, which leads to a basis-dependent problem tackled by a quantum coin method \cite{lucamarini2015}. This affects the evaluation of the phase error rate
\begin{equation}
	\label{eq:phase_error}
	\begin{aligned}
		e_{1}^X = &~ e_\text{1,bit}^X+4\Delta^{\prime}(1-\Delta^{\prime})(1-2e_\text{1,bit}^X)\\
		& +4(1-2\Delta^{\prime})\sqrt{\Delta^{\prime}(1-\Delta^{\prime})e_\text{1,bit}^X(1-e_\text{1,bit}^X)},
	\end{aligned}
\end{equation}
where $e_\text{1,bit}^X$ is the bit error rate and $\Delta^{\prime}$ is used to quantify the basis-dependence.

The leakage of the intensity-encoding information leads to a distinguishable-decoy-states problem, treated with a trace distance method \cite{tamaki2016,sun2021}. It introduces a deviation in the parameter estimation in the decoy-state analysis
\begin{equation}
	\label{eq:Y_n}
	\begin{split}
		&\left|Y_{n}^{j}-\left[q_{nkl}Y_{n}^{k}+\left(1-q_{nkl}\right)Y_{n}^{l}\right]\right|\leq D_{n,j,k,l}, \\
		&\left|Y_{n}^{j}e_{n}^{j}-\left[q_{nkl}Y_{n}^{k}e_{n}^{k}+\left(1-q_{nkl}\right)Y_{n}^{l}e_{n}^{l}\right]\right|\leq D_{n,j,k,l},
	\end{split}
\end{equation}
where $Y_{n}^{j}$ ($e_{n}^{j}$) is the yield (error rate) of $n$-photon signals under the THA given the intensities selected by Alice $j\in \{\mu,\nu,\omega\}$, $D_{n,j,k,l}$ is the trace distance, and $q_{nkl}$ is the conditional probability to have selected the intensity setting $k$ given that the pulse contains $n$ photons.

\begin{figure*}
	\includegraphics{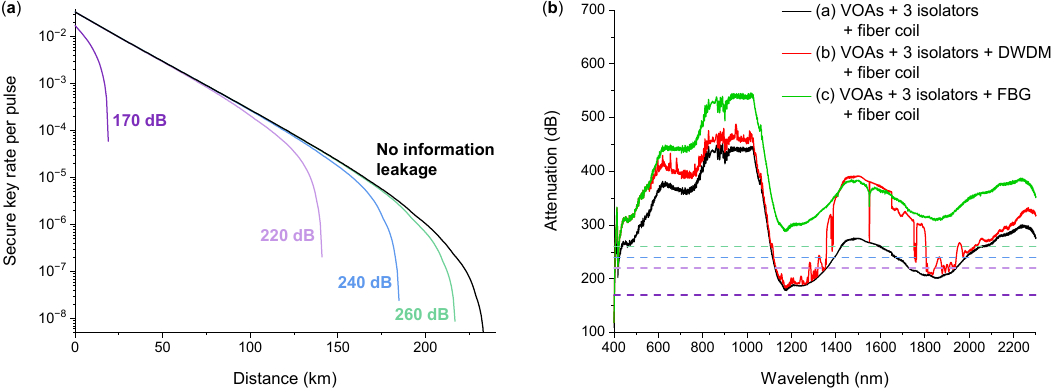}
	\caption{Security analysis of BB84 QKD under the THA. (a)~Secure key rate for different attenuations of Trojan photons. (b)~Total measured attenuation of Trojan photons for the three source configurations shown in \cref{fig:source_configurations} under the assumption of unity backreflection.}
	\label{fig:BB84_key_rate_dB}
\end{figure*}

We investigate the secure key rate of a decoy-state BB84 protocol \cite{ma2005} under the THA based on \cite{lucamarini2015,tamaki2016}. Its lower bound can be written as
\begin{equation}
	R = \left\{\left(\mu e^{-\mu}\right)Y_{1}^{Z,L}\left[1-h_2\left(e_{1}^{X,U}\right)\right]-f_eQ_{Z}^{\mu}h_2\left(E_{Z}^{\mu}\right)\right\},
\end{equation}
where $Q_{Z}^{\mu}$ and $E_{Z}^{\mu}$ are the gain and QBER in $Z$ basis; $Y_{1}^{Z,L}$ ($e_{1}^{X,U}$) is the lower (upper) bound of the single-photon yield (error rate) in $Z$ basis ($X$ basis); $f_e$ is the error correction inefficiency; $h_2\left(x\right)$ is the binary entropy function. The key to calculating the secure key rate is to accurately estimate the single-photon parameters $Y_{1}^{Z,L}$ and $e_{1}^{X,U}$. We use a decoy analysis method from \cite{ma2005}. By introducing \cref{eq:phase_error,eq:Y_n} into the calculation of the secure key rate, we account for the THA.

We also investigate a decoy-state measurement-device-independent (MDI) QKD protocol \cite{lo2012}. Its secure key rate can be written as \cite{zhou2016,wang2019}
\begin{equation}
	\label{eq:key_rate_formula}
	\begin{aligned}
		R = p_{s_A} p_{s_B} \Big\{ \!& s_A e^{-s_A} s_B e^{-s_B} Y_{11}^{X,L}\left[1-h_2\left(e_{11}^{X,U}\right)\right] \\
		& -f_e Q_{Z}^{ss} h_2 \left(E_{Z}^{ss}\right) \Big\},
	\end{aligned}
\end{equation}
where $p_{s_A}$ and $p_{s_B}$ are the probabilities of Alice and Bob sending the signal states, $Q_{Z}^{ss}$ and $E_{Z}^{ss}$ are the gain and QBER in $Z$ basis, $Y_{11}^{X,L}$ ($e_{11}^{X,U}$) is the lower (upper) bound of the single-photon yield (error rate) in $X$ basis, and $s_A$ and $s_B$ are the intensities of signal states. We use a decoy analysis method from \cite{xu2013}. By introducing \cref{eq:phase_error,eq:Y_n} into the calculation of the secure key rate, we account for the THA.

The maximum power that can be injected within $400$--$2300$~\si{\nano\meter} is not known precisely. Here we adopt the value of 15.6~\si{\watt} at 2300~\si{\nano\meter} \cite{lucamarini2015}. Although this might be a slight underestimate of fiber power-carrying capability, it is in line with power levels observed to cause destructive effects in components and in the fiber \cite{kashyap1988,davis1997,huang2020,ponosova2022}, and therefore is a plausible order-of-magnitude estimate of Eve's limit. For simplicity, we set $N = 1.9 \times 10^{20}$~photon/s.

\medskip
\emph{Prepare-and-measure BB84 QKD system.}
We calculate the secure key rate of the BB84 QKD system as detailed above, using system parameters listed in \cref{tab:system_parameter} and protocol specification detailed in Appendix~\labelcref{sec:specs-protocol}. \Cref{fig:BB84_key_rate_dB}(a) shows the key rate for $\gamma = 0$ (no information leakage) and four different attenuations of Trojan photons, resulting in reduction of the maximum key generation distance to $93\%$, $79\%$, $60\%$, and $8\%$.

\begin{table}[b]
	\caption{Experimental parameters \cite{tan2021,li2021} we use in the calculation of secure key rate. The previous QKD systems \cite{tan2021,li2021} had a detector efficiency of $49.5\%$ and receiver insertion loss of $1.1~\si{\deci\bel}$, resulting in the total receiver detector efficiency of $38\%$ we use here.}
	\label{tab:system_parameter}
	\begin{tblr}[t]{llc}
		\hline\hline
		Channel loss coefficient & $\alpha$~(dB/km) & $0.2$ \\ 
		Clock rate & $f$~(Hz) & $1.25 \times 10^{9}$ \\
		Background rate & $Y_0$ & $8 \times 10^{-8}$ \\
		Total misalignment error & $e_d$ & $2\%$ \\
		Detector efficiency & $\eta_\text{det}$ & $38\%$ \\
		Error correction inefficiency & $f_e$ & $1.16$ \\
		\hline\hline
	\end{tblr}
\end{table}

The actual measured attenuation of Trojan photons, calculated via \cref{eq:configuration-in-out-simplified}, is shown in \cref{fig:BB84_key_rate_dB}(b). The driving voltages of the two VOA samples are set here to $9$ and $10~\si{\volt}$, corresponding to attenuation of about $13$ and $18~\si{\deci\bel}$ at $1550~\si{\nano\meter}$. Their resulting attenuation of the order of $30~\si{\deci\bel}$ is typical for a gigahertz-rate QKD system's source \cite{li2023}. These plots show that the source configuration (c) enables the QKD system to achieve over $93\%$ maximum key generation distance under the THA over the entire wavelength range. However, configurations (a) and (b) both have two significant high-risk wavelength bands near $1200$ and $1900~\si{\nano\meter}$, where the guaranteed key generation distance drops well below $60\%$. Lower attenuation in these bands stems from the deficiencies in spectral characteristics of the isolator, VOA, and DWDM. The fiber Bragg grating-based filter in the configuration (c) efficiently compensates for these.

\begin{figure*}
	\includegraphics{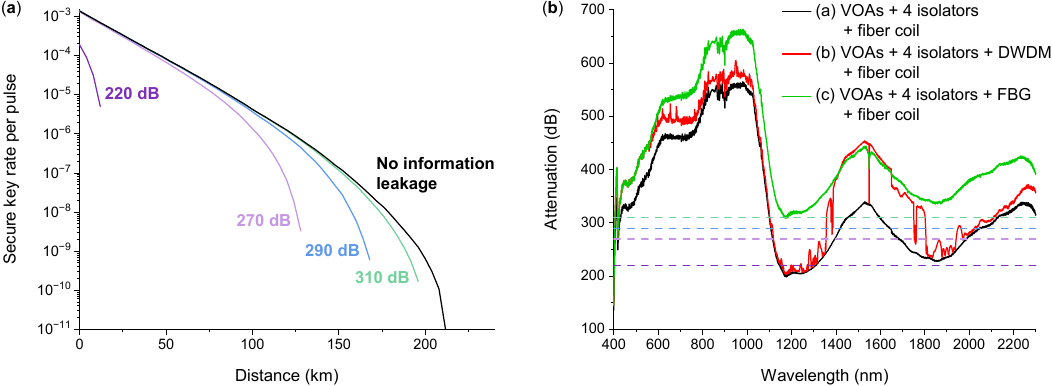}
	\caption{Security analysis of MDI QKD under the THA. (a)~Secure key rate for different attenuations of Trojan photons. (b)~Total measured attenuation of Trojan photons for the three source configurations shown in \cref{fig:source_configurations} with four isolators in each instead of three. We conservatively assume unity backreflection.}
	\label{fig:MDI_key_rate_dB}
\end{figure*}

\medskip
\emph{MDI-QKD system.}
Measurement-device-independent QKD is immune to detection-side attacks. Both Alice and Bob function as transmitters, encoding standard BB84 quantum states and sending them to an untrusted relay Charlie, where Bell-state measurements are performed. The security of the transmitter thus becomes vital. In our analysis, we assume Eve attacks both transmitters of the system simultaneously. We use system parameters listed in \cref{tab:system_parameter} and protocol specification detailed in Appendix~\labelcref{sec:specs-protocol}. \Cref{fig:MDI_key_rate_dB}(a) shows the key rate for $\gamma = 0$ (no information leakage) and four different attenuations of Trojan photons, resulting in reduction of the maximum key generation distance to $93\%$, $79\%$, $60\%$, and $6\%$.

The actual measured attenuation of Trojan photons is shown in \cref{fig:MDI_key_rate_dB}(b). Since this system requires higher attenuation, we have increased the number of isolators in all the source configurations to four. The settings of VOAs are the same as those in the BB84 QKD system, because the source's function is similar. Similarly to the BB84 QKD system, the source configuration (c) allows $93\%$ maximum key generation distance under the THA over the entire wavelength range. Configurations (a) and (b) fail to guarantee key generation, owing to the high-risk band near $1200~\si{\nano\meter}$.

We consider two experimental MDI-QKD systems, a free-space one \cite{cao2020} and fiber-optic one \cite{wei2020}. In both systems, the source is implemented with fiber optics. In the free-space system \cite{cao2020}, there are two components between the modulators and the transmitter telescope: one VOA and one circulator connected via its ports $1 \rightarrow 2$. These two components provide less than $70~\si{\decibel}$ total attenuation of Trojan photons at $2000$--$2200~\si{\nano\meter}$. This is lower than the threshold estimated in \cref{fig:MDI_key_rate_dB}(a). The fiber-optic system \cite{wei2020} implements the intensity modulator, attenuators, and polarisation modulator in an integrated-optics Si chip, followed by a single fiber-optic isolator before the communication channel. Although the integrated chip has the advantage of lower reflectivity than bulk optics, one isolator is not sufficient \cite{tan2021}. Therefore, both experimental systems need additional isolating and filtering components for the protection against THA in a rigorous manner.

\subsection{Induced-photorefraction attack}

In this attack, Eve injects short-wavelength light into QKD device that creates a space charge in its lithium-niobate electrooptic modulators and thus changes their photorefractive properties \cite{ye2023,lu2023,han2023}. Illumination power at the modulator of $3~\si{\nano\watt}$ ($400~\si{\micro\watt}$) at $405~\si{\nano\meter}$ ($532~\si{\nano\meter}$) is sufficient to induce measurable changes in its characteristics.

\begin{figure}
	\includegraphics{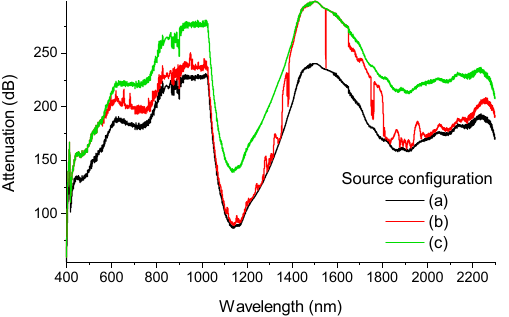}
	\caption{Total measured attenuation of Eve's light reaching the modulators in the three source configurations shown in \cref{fig:source_configurations}.}
	\label{fig:induced-photorefraction-dB}
\end{figure}

We can easily calculate $\gamma(\lambda)$ for the attack channel via \cref{eq:configuration-in} in our source configurations (\cref{fig:induced-photorefraction-dB}) and upper-bound Eve's power reaching the modulators. However, the modulators' susceptibility has only been measured at the two discrete wavelengths. A general security proof that takes this imperfection into account is not available. With this limited knowledge, only a preliminary estimate of the system's security can be made.

Assuming the maximum cw power that Eve can inject into the fiber to be $15.6~\si{\watt}$ just like in the Trojan-horse attack, the power that reaches the modulators in configuration (a) at $405~\si{\nano\meter}$ ($532~\si{\nano\meter}$) is $3.3 \times 10^{-10}~\si{\watt}$ ($7.2 \times 10^{-15}~\si{\watt}$). This is 1 (11) orders of magnitude below that causing measurable effects \cite{ye2023,han2023}. Source configurations (b) and (c) provide slightly higher attenuation. The safety margin at $405~\si{\nano\meter}$ is small and the attenuation drops rapidly towards shorter wavelengths (\cref{fig:induced-photorefraction-dB}), preventing us from drawing a reliable conclusion. If the short-wavelength suppression physical filter is implemented (\cref{sec:physical_filter}), the sources can probably be considered safe. However, we stress that the general security proof providing the secure key rate $R$ is needed to confirm this and the modulator susceptibility $S(\lambda)$ needs to be characterised at other wavelengths. For the latter characterisation, our testbench may be supplemented with a monochromator \cite{makarov2024}.

\subsection{Detector-backflash attack}

This is a passive attack, in which Eve measures the polarisation or other properties of light emitted into the channel by QKD receiver device \cite{meda2017,pinheiro2018,koehler-sidki2020}. The spectrally broad emission originates in avalanche photodiodes and may identify which detector has clicked, thus leaking key information.

In a typical QKD receiver, backflash light passes several components on its way to the communication channel, such as DWDM and isolators similar to those we have characterised \cite{makarov2024}. The transmission of attack channel $\gamma(\lambda)$ is calculated by \cref{eq:configuration-out}. While security proofs providing $R$ in the presence of this attack are available \cite{pinheiro2018,koehler-sidki2020,molotkov2020}, they require as input the probability of photon emission from the receiver $\mu_\text{leak} = \int_\lambda \gamma(\lambda) S(\lambda) \,d\lambda$, where $S(\lambda)$ is the probability density of photon emission from the detector. The latter spectral distribution is challenging to measure, owing to high sensitivity required. The spectrometers in our testbench lack it. The characterisation has to be done with a set of spectral filters or tunable filter followed by a single-photon detector \cite{marini2017,shi2017,meda2017}.

\section{Conclusion}
\label{sec:conclusion}

We have demonstrated a characterisation methodology to achieve full-spectrum security of QKD systems. Since all known implementations of QKD utilise photons as information carriers, it should be applicable to most optical attacks on them. The full analysis of each attack needs data on wavelength-dependent transmission of the attack channel, target susceptibility, and security proof that accounts for these.

We have used our testbench to test passive fiber-optic components in a wide spectral range of $400$ to $2300~\si{\nano\meter}$. Our experimental results confirm that the optical device characteristics deviate significantly at different wavelengths, which may lead to potential security vulnerabilities. We have evaluated the protection of QKD sources against THA and identified that the source configuration (c) with the FBG-based filter guarantees the security of typical one-way decoy-state BB84 and MDI-QKD systems. We have outlined how to apply our methodology to induced-photorefraction and detector-backflash attacks.

Our testbench and characterisation methodology can be used for certification of QKD systems against other attacks as well. This promotes secure implementation of QKD and its standardisation.

\begin{acknowledgments}

We thank Anqi Huang for discussions and her help with this study.

\emph{Funding:}
%This work was funded by the
National Key Research and Development (R\&D) Plan of China (grant 2020YFA0309701), National Natural Science Foundation of China (grant 62031024), Innovation Program for Quantum Science and Technology (grant 2021ZD0300300), Shanghai Municipal Science and Technology Major Project (grant 2019SHZDZX01), Shanghai Academic/Technology Research Leader (21XD1403800), Shanghai Science and Technology Development Funds (22JC1402900), Key-Area Research and Development Program of Guangdong Province (2020B0303020001), Anhui Initiative in Quantum Information Technologies, and Chinese Academy of Sciences. M.P.\ and V.M.\ acknowledge funding from the Galician Regional Government (consolidation of research units: atlanTTic and own funding through the ``Planes Complementarios de I+D+I con las Comunidades Autonomas'' in Quantum Communication), MICIN with funding from the European Union NextGenerationEU (PRTR-C17.I1), the ``Hub Nacional de Excelencia en Comunicaciones Cu{\' a}nticas'' funded by the Spanish Ministry for Digital Transformation and the Public Service and the European Union NextGenerationEU, and the European Union's Horizon Europe Framework Programme under Marie Sk\l{}odowska-Curie grant 101072637 (project QSI) and project ``Quantum Security Networks Partnership'' (QSNP; grant 101114043). F.X.\ acknowledges support from the New Cornerstone Science Foundation through the Xplorer Prize.

%\medskip \emph{Author contributions:} H.T.\ performed the experiments and analysed the data. M.P.\ designed the testbench and analysed the data. V.M.,\ W.Z.,\ and L.H.\ analysed the data. H.T.\ and V.M.\ wrote the article with input from all authors. S.-K.L.,\ V.M.,\ and F.X.\ supervised the project.

\end{acknowledgments}

\appendix

\section{QKD protocol specification}
\label{sec:specs-protocol}

In the security analysis of the BB84 QKD system, we study the three-intensity decoy state BB84 QKD protocol \cite{ma2005}. Alice chooses her bit value uniformly at random. Then, the bases $Z$ and $X$ are selected with probabilities $P_Z$ and $1-P_Z$ and the secret key is extracted from the events whereby Alice and Bob both chose the $Z$ basis. Each pulse is randomly prepared in one of the three intensities \{$\mu$, $\nu$, $\omega$\} chosen with probabilities $P_{\mu}$, $P_{\nu}$, and $(1-P_{\mu}-P_{\nu})$. The intensities satisfy $\mu > \nu + \omega $ and $ \mu > \nu > \omega$. Here, $\mu$ denotes the intensity of the signal state.  We perform a full optimisation of parameters \{$\mu$, $\nu$, $P_{\mu}$, $P_{\nu}$, $P_Z$\}. In parameter estimation, we use the analytical approaches \cite{ma2005,huang2018}, and consider statistical fluctuations \cite{lim2014} and deviations caused by THA \cite{lucamarini2015,tamaki2016}.

In the security analysis of the MDI-QKD system, we study the four-intensity decoy state MDI-QKD protocol \cite{zhou2016,wang2019}. There are three intensities \{$\mu$, $\nu$, $\omega$\} in the $X$ basis for the decoy-state analysis and one signal intensity \{$s$\} in the $Z$ basis for secret key generation. We consider a symmetric channel loss where Alice and Bob use the same parameters. Including the probabilities $P$ for each intensity, both Alice and Bob use the same group of six parameters \{$s$, $\mu$, $\nu$, $P_s$, $P_{\mu}$, $P_{\nu}$\}. We perform a full optimisation of parameters. In parameter estimation, we use the analytical approaches \cite{xu2013,tan2021}, and consider statistical fluctuations \cite{wang2019} and deviations caused by THA \cite{tan2021}.

\def\bibsection{\medskip\begin{center}\rule{0.5\columnwidth}{.8pt}\end{center}\medskip} % Redefines bibliography separator to single-column. This reduces chances of float placement bugs in the last page.
%\bibliography{library}

\end{document}